\documentclass[a4paper]{article}

\usepackage{INTERSPEECH2022}
\usepackage{multirow}
\usepackage{bm}
\usepackage{amssymb,amsmath,graphicx,bbm,color,booktabs}
\usepackage{amsopn}

\newcommand{\vm}{\bm{m}}

\newcommand{\vp}{\bm{p}}

\newcommand{\vx}{\bm{x}}               
\newcommand{\vy}{\bm{y}}

\newcommand{\R}{\mathbb{R}}

\newcommand{\tT}{\tau}

\newcommand{\sx}{\bar{\vx}}

\newcommand{\tx}{\bar{\vx}_k^{\tau}}

\newcommand{\tp}{\bar{\vp}}

\newcommand{\ty}{\bar{\vy}_k^{\tau}}
\newcommand{\tm}{\bar{\vm}}

\newcommand{\x}{\vx}

\newcommand{\sy}{\bar{\vy}}
\renewcommand{\sp}{\bar{p}}
\newcommand{\st}{\bar{t}}

\renewcommand{\eqref}[1]{Eq.~(\ref{#1})}

\usepackage{url}

\title{Correcting Mispronunciations in Speech using Spectrogram Inpainting}
\name{Talia Ben-Simon$^1$, Felix Kreuk$^1$, Faten Awwad$^2$, Jacob T. Cohen$^2$, Joseph Keshet$^3$}
\address{
  $^1$Department of Computer Science, Bar-Ilan University, Ramat-Gan, Israel\\
  $^2$Department of Otolaryngology Head and Neck Surgery, Rambam Health Care Campus, Israel\\
  $^3$Facullty of Electrical and Computer Engineering, Technion--Israel Institute of Technology, Israel}%
\email{jkeshet@technion.ac.il}

\begin{document}

\maketitle
\begin{abstract}
Learning a new language involves constantly comparing speech productions with reference productions from the environment. Early in speech acquisition, children make articulatory adjustments to match their caregivers' speech. Grownup learners of a language tweak their speech to match the tutor reference. This paper proposes a method to synthetically generate correct pronunciation feedback given incorrect production. Furthermore, our aim is to generate the corrected production while maintaining the speaker's original voice. 

The system prompts the user to pronounce a phrase. The speech is recorded, and the samples associated with the inaccurate phoneme are masked with zeros. This waveform serves as an input to a speech generator, implemented as a deep learning inpainting system with a U-net architecture, and trained to output a reconstructed speech. The training set is composed of unimpaired proper speech examples, and the generator is trained to reconstruct the original proper speech. We evaluated the performance of our system on phoneme replacement of minimal pair words of English as well as on children with pronunciation disorders. Results suggest that human listeners slightly prefer our generated speech over a smoothed replacement of the inaccurate phoneme with a production of a different speaker. 
\end{abstract}


\section{Introduction}

Learning a language involves constantly comparing one's own speech production to reference examples given by the speakers of the language. Children at the early stages of language acquisition make articulatory adjustments to match better the reference models, and grownups learning a new language adjust their speech so as it becomes more similar to their teachers.

Children and adults who suffer from speech sound disorder (SSD) have persistent difficulty pronouncing words or sounds correctly. They are often referred to speech therapy, where the work is focused on learning the motor skills of sound production by providing feedback in the form of the correct production. The most critical part of the treatment is the feedback, which helps develop a proper model of pronunciation \cite{strombergsson2014children}.

In this paper, we would like to propose an algorithm to generate synthetic feedback of proper pronunciation from the wrong pronunciation. Assume, for example, that Ron has a problem pronouncing /r/ in American English. It can be because he suffers from SSD or since his native language is Greek and he is trying to learn English. The system will ask Ron to pronounce the word ``\emph{right}.'' Ron will say ``\emph{white},'' mispronouncing /r/ with /w/. The system will say to Ron: ``You pronounced \emph{white}, but you should have pronounced \emph{right}'' -- where the word ``\emph{right}'' will be synthesized in Ron's own voice. Ron will be able to compare his production with the desired production and try to fix it.

Our goal is to automatically synthesize a waveform with proper pronunciation generated from the waveform of wrong pronunciation, which will serve as a reference or feedback for the speaker. The generated speech should maintain the original voice of the speaker, as it was found to be a critical factor in providing effective feedback \cite{strombergsson2014children}.

The most straightforward approach to this problem is to learn a mapping between the wrongly pronounced phoneme to a correct production of the phoneme. To do so, we need a collection of parallel aligned data of the same speaker. That is a recording of the child's voice pronouncing the same word wrongly and correctly. Unfortunately, it is impractical to collect such data, and therefore we are focused on designing an algorithm that can handle non-parallel data.

Str{\"o}mbergsson \cite{strombergsson2011segmental, strombergsson2012synthetic, strombergsson2014children, strombergsson2015acoustic} was the first to suggest a method for automatic re-synthesis of recorded wrongly-pronounced children's speech. In that line of work, a unit selection approach was used to replace an initial voiceless plosive in a recorded speech sample with another voiceless plosive, retrieved from a recording of another child. The method was evaluated with adult listeners, and preliminary results from an evaluation with children with typical speech as speakers/listeners were reported \cite{strombergsson2014children}. While this technique allows previously impracticable exploration of self-monitoring skills in children producing misarticulated speech, e.g., velar fronting, it is an \emph{ad-hoc} method, which is limited to the correction of utterances with incorrect production of voiceless plosives. Our goal, however, is to be able to modify any phoneme while maintaining the speaker's voice.

Recently, Biadsy \emph{et al.} \cite{biadsy2019parrotron} proposed an end-to-end speech-to-speech conversion model called \emph{Parrotron} that maps an input spectrogram to a target spectrogram, without utilizing any intermediate discrete representation. The authors showed that the model could be trained to normalize speech from any speaker, regardless of accent, prosody, and background noise, into the voice of a single canonical target speaker with a fixed accent and consistent articulation and prosody. This model is not suitable for our task. Firstly, there is no parallel data for our task (Paratron is trained on 30,000 hours of generic synthetically generated speech). Secondly, the \emph{Parrotron} model is designed to produce a generic speech, while in our task, it is critical to retain the child's original voice. Similarly, algorithms for foreign accent conversion \cite{zhao19f_interspeech} and pronunciation prediction \cite{soboleva2021replacing} will not meet our lack of data and the goal of maintaining the speaker voice.



Methods based on Automatic speech recognition (ASR) have been widely used to improve and assist speech and language therapy in recent years \cite{mckechnie2018automated, furlong2017computer, popovici2012professional}. Recent work provides feedback through games to encourage children's engagement in otherwise repetitive tasks while providing an indirect progress report for evaluation \cite{shahin2015tabby, navarro2014talking}. The feedback may be a real-time response like a simple score or visual traffic light concept response or, in other cases, an immediate real-time impact on the user's progress in the game with remote clinical interference like updating task \cite{ahmed2018speech}. This type of feedback does not actively suggest to the users how to improve their pronunciation, and they are just noted if their production is correct or not. More effective feedback was proposed by \cite{saz2009tools}. The children are shown a form of the formants, and interactively they need to reach a target goal \cite{saz2009tools}. 

Our work is related to inpainting works in vision and speech. The concept matched our goal by removing a segment while retaining the surrounding information for keeping the speaker's identity. Our base model architecture is inspired from \cite{liu2018image}. There have been several studies on inpainting, particularly on speech, that strengthen our concept of context information with inpainting \cite{marafioti2019context, perraudin2018inpainting, kreme2018phase, toumi2019audio, mokry2019introducing, chang2019deep, zhou2019vision, kegler2019deep}.

Although inpainting provides a suitable \emph{context encoder}, it can not change significant components, let alone correct speech segments without any guidance. 

This paper represents an inpainting method to correct speech while retaining the speaker's voice. 
Samples of our model can be found on the project website \url{https://github.com/MLSpeech/spoken_feedback}.

\section{Problem Setting}\label{sec:problem_setting}

Recall the input speech is generated by prompting the user, who potentially has a production error, to pronounce a word (or a phrase). Hence, the canonical target word is known, and we are aware that it is most likely different from the actual surface pronunciation. Throughout the paper, we will assume that only a single phoneme is mispronounced\footnote{If the speaker mispronounces several phonemes, the chosen target word sequence can be designed to include only a single mispronounced phoneme at a time.}.

Formally, let $\sx = (\x_1,\ldots,\x_T)$ denote the input speech utterance, where each $\x_t\in\R^D$ is a $D$-dimensional vector for $1\leq t \leq T$. In this paper, this vector is composed of the frequency bins of the Mel-spectrum, but other representations could be used as well. The duration of the speech utterance, $T$, is not fixed. We denote by $\sx_{t_1}^{t_2} = (\vx_{t_1},\ldots,\vx_{t_2-1})$ the portion of the speech utterance from time frame $t_1$ to time frame $t_2-1$.

The desired \emph{correct} phoneme sequence corresponding to $\sx$ is denoted as $\sp=(p_1,\ldots,p_K)$, where $p_k\in\mathcal{P}$, and $\mathcal{P}$ is the set of the phonemes in the language. We denote the segmentation of the phonemes in the speech by $\st = (t_1, \ldots, t_K)$, where $1 \le t_k < T$ is the start time (in number of frames) of the phoneme $p_k$. The speech segment associated with the $k$-th phoneme $p_k$ is denoted as $\sx_{t_k}^{t_{k+1}}$. 

Phoneme alignment and segmentation can be obtained by using a forced-alignment. In our experiments we found out that applying force-alignment with the canonical phoneme sequence and the actual mispronounced speech, gave good-enough segmentation for the correction process. Note that practically, we care only for the phonemes in the vicinity of the mispronounced phoneme.

Denote by $\rho^*$ the phoneme symbol that is wrongly pronounced in $\sx$ and needs to be corrected. Without loss of generality, let the $k$-th phoneme in this sequence be the mispronounced phoneme, namely, $p_k=\rho^*$. Our goal is to generate a new speech utterance $\sy$ that is based on the input speech utterance $\sx$, where the segment $\sx_{t_k}^{t_{k+1}}$ of the wrongly-pronounced phoneme is replaced with a newly generated segment $\sy_{t_k}^{t_{k+1}}$ representing the \emph{desired pronunciation} of the phoneme $\rho^*$. The replaced phoneme might not necessarily be of the same duration as the originally pronounced phoneme, and then $t_k$ and the following time segments are changed accordingly. 

\section{Method}\label{sec:method}



Our method first replaces the mispronounced phoneme with complete silence, i.e., the samples that correspond to this phoneme are replaced with zeros. Then a neural network called \emph{generator} gets this processed speech as input and is aimed at reconstructing back the original speech waveform. During the training the generator works on examples of correctly pronounced speech segments, where a segment associated with a phoneme $\rho^*$ is replaced with zeros. Its goal is to regenerate the missing speech. At inference time, the generator is provided with misproducted speech, and it generates a correct speech, since it was trained only on examples of correctly pronounced speech.

The generator works on input speech segments of a fixed duration. Let us denote by $G$ the generator that gets as input a speech segment of duration $\tT$. The output of the generator is a speech segment of the same duration $\tT$. We set this duration to be longer by 30\% than the maximum phoneme $\rho^*$ duration so as to have enough information of the speaker's voice in the replacement context. The speech segment corresponding to the phoneme $p_k$ with its neighborhood is denoted by $\tx$. 

Define $\bar{\vm}_{t_1}^{t_2}$ a binary masking vector of length $\tT$ whose elements are either $0$ or $1$. This vector is set to zeros for frames in the region $[t_1, t_{2}-1]$ and to ones otherwise. Specifically, we denote $\tm_k=\bar{\vm}_{t_k}^{t_{k+1}}$ to be the mask vector of phoneme $k$. Similarly, we define $\tp_k=\bar{\vp}_{t_k}^{t_{k+1}}$ the phoneme embedding sequence at the same region.

The input to the generator is the speech segment when the mask is applied to it, $\tm_k \odot \tx$, where $\odot$ stands for the element-by-element product, and a sequence of phoneme embedding of the same length $\tp_k$. The generator's output, $\ty = G(\tx, \tp_k)$, is a sequence of vectors with the same representation as $\tx$ (Mel-spectrum bins), and with the same duration. Importantly, the generator uses the portions of the speech which was not set to zeros to recover the missing phoneme and maintain the speech smoothness. We now turn to describe how the generator $G$ is trained effectively. 


\subsection{Training objectives}

The most straightforward way to train $G$ is simply by letting it recover the acoustic signal spanned by the mispronounced phoneme $p_k$ and its neighborhood, including the previous and the following phonemes. Our motivation is to keep as much of the original signal and fill in the ``missing'' phoneme, which is replaced with zeros. The basic model is based on encoder-decoder-like architecture. We aim to train an encoder with a long-enough receptive field that captures the contextual information of the phoneme's vicinity and a decoder that will be able to generate the missing phoneme while preserving its surrounding accurately.


The model is trained on unimpaired speech. Specifically, we mask the speech corresponding to the phoneme $p_k$ and train $G$ to recover it. In the basic model we would like to fulfill two complementary objectives: (i) the recovered phoneme $p_k$ should be similar to the desired phoneme $\rho^*$, which in this case is the original phoneme prior to its masking; and (ii) the neighborhood of the phoneme should be similar to its original neighborhood. 

The first objective deals with the recovered phoneme. The output of the generator $\ty$ is a speech signal that spans the recovered phoneme as well as its environment. The span area of the recovered phoneme is $(1-\tm_k) \odot \ty$ and it should be similar to the original signal $(1-\tm_k) \odot \tx$. Similarly, the second objective deals with the phoneme neighborhood, where the recovered neighborhood $\tm_k \odot \ty$ should be similar to the original environment $\tm_k \odot \tx$. In our basic setting we measure similarity using the $\ell_1$ loss function, and the overall objective is:
\begin{multline}\label{eq:objective_basic}
\lambda_1 \Big| 
(1-\tm_k) \odot \ty -  (1-\tm_k) \odot \tx \Big| \\~+~ 
\lambda_2  \Big| \tm_k \odot \ty ~,~  \tm_k \odot \tx \Big|
\end{multline}





The generator was implemented as a Convolutional Neural Network (CNN) in a U-net architecture \cite{ronneberger2015u}. 
Similar architecture was used in the task of restoring missing speech as illustrated in \cite{marafioti2019context,chang2019deep}. It seems, however, that this objective and architecture is less suitable to our task. Empirically, our system was able to recover the speech of a speaker that was not seen in the training set. However, when the speaker had speech impairment, the recovered speech was still impaired, and the mispronounced phoneme was recovered instead of the desired phoneme $\rho^*$. The reason is that the objective in \eqref{eq:objective_basic} is motivated to recover speech in general and not the desired phoneme $\rho^*$ in particular. In the next subsections we will elaborate on our method of improving the basic model.

\begin{table*}
\small
\renewcommand{\arraystretch}{0.9}
  \centering
    \caption{Relative MOS gap - negative values indicate higher score for reference audio}
    \vspace{-0.2cm}
  \label{tab:phonemes_conditional_mos}
  \begin{tabular}{l |l|ccccc}
    \toprule
    {Audio type}&
    {Reference}  & {MOS GAP} &
    {$\rho^*=S$ }  & {$\rho^*=SH$}  & {$\rho^*=W$ }  & {$\rho^*=R$ }\\
        \midrule
       \textit{Vocoder only}& \textit{Original}& \textbf{-0.045 $\pm$ 0.59}& \text{-0.07 $\pm$ 0.50}& \text{0.01 $\pm$ 0.66} & \text{-0.03 $\pm$ 0.61}& \text{-0.09 $\pm$ 0.53}\\
       \textit{Generated}& \textit{Vocoder only}& \textbf{-0.219 $\pm$ 0.64} & \text{-0.22 $\pm$ 0.67} & \text{-0.05 $\pm$ 0.60}& \text{-0.24 $\pm$ 0.59}& \text{-0.28 $\pm$ 0.68}\\
       \textit{Smooth concat.}& \textit{Generated}& \textbf{-0.41 $\pm$ 0.80}& \text{-0.11 $\pm$ 0.77}& \text{-0.07 $\pm$ 0.73}& \textbf{-0.75 $\pm$ 0.72}& \textbf{-0.55 $\pm$ 0.79}\\
    \bottomrule
  \end{tabular}
  \vspace{-0.5cm}
\end{table*}

\subsection{Acoustic embedding}

To further steer the generator to generate the correct phoneme rather than reconstruct the mispronounced speech we used phoneme acoustic embedding. While previously the phoneme embedding was created based on the phoneme symbol, now we propose to use embedding which is based on the acoustic characteristic of each  phoneme. The acoustic phoneme embedding allows the model to measure whether two acoustic inputs can be associated with the same phoneme or not. 




We trained a Siamese structure with cosine similarity loss applied between the embedding vectors. Siamese network uses the same weights while working in tandem on two different input vectors to compute comparable output vectors. To the overall objective of our generator, we add two loss terms. First loss is aimed at minimizing the cosine distance between the acoustic embedding of the generated phoneme $\rho^*$ and random samples of the same phoneme. For the second loss term, we picked a contrastive phoneme $q$ and generated a speech with this phoneme (and not $\rho^*$). The loss is aimed at minimizing the cosine distance between the acoustic embedding of speech generated with contrastive phoneme and random samples of that phoneme.

\subsection{Inference}
At inference time, we transform sample $\tx$ by feeding it to the generator $G$ along with the target phoneme embeddings $\tp_k$. The new sample $\ty$ is given by $G(\tm_k \odot \tx, \tp_k)$.


While $G$ generates the desired spectrogram features, we are missing phase information to invert back to audio. We examined a few possibilities to get the corresponding phase. Although using the initial wave phase gave us the best results on the surrounding area, it gave poor results on a different generated phoneme. Our next approach was using the Griffin-Lim algorithm to estimate the missing phase information, which still fell a little short. The results with WaveGlow \cite{prenger2019waveglow}, a flow-based model that generates speech given Mel spectrograms, were significantly better but some baseline noise was undesirably added to the audio. We got the best results with HiFi-GAN and used it to produce the final waveform \cite{kong2020hifi}. An additional considerable advantage of the HiFi-GAN compared to other methods is in training simplicity and inference speed.

\section{Empirical Evaluation}\label{sec:results}
Our generator $G$ is composed of a U-net encoder-decoder. We trained our model for 450 epochs using Adam optimizer, a batch size of 100, an initial learning rate of $10^{- 4}$ and early stopping. Each component includes 5 1D-convolutional layers with $p$-ReLU activation, in which 2 layers are down-sampling/up-sampling layers respectively achieved with kernels of size 3, stride 2. No batch norm was used since it showed no improvement or even resulted in a slight decrease.

The Siamese network for the acoustic embedding  consisted of one layer of bidirectional GRU with hidden size of 300, followed by a linear layer with ReLU.

We fine-tuned HiFi-GAN model with generated examples on LibriSpeech dataset. The examples were generated by taking a sample $\tx$ and feeding it to $G$ along with the original target phonemes $\tp_k$ and set $\tm_k$ the mask vector to all $1$. The new sample is given by $G(\tm \odot \tx, \tp_k)$ as the spectogram matching the original audio clip for fine-tuning the model.

All our models were trained on TIMIT dataset after they were resampled to 22.5 kHz (to match HiFi-GAN). We used the original train and test split, and further split the training corpus to train and validation sets with a ratio of 0.8 and 0.2, respectively. We used the Mel-spectrogram of the original audio as the input, with 80 bins, where each bin is normalized by the filter length. The Mel-spectrogram parameters were FFT size 1024, hop size 256, and window size 1024.

\subsection{Results}

This section will present our results and the testing methods we used.
We provide an evaluation on two datasets: (i) partial utterances from LibriSpeech-clean-100 aligned with Montreal Forced Aligner \cite{mcauliffe2017montreal}: mimicking speech correction with minimal-pair words since mispronunciation usually occurs between specific phonemes in every language, for example, \emph{run, one},\emph{ship, sip}; and (ii) audio recordings of patients diagnosed with SSD in Hebrew. We evaluated our approach with qualitative experiments with human listeners using a crowd-sourcing platform on both datasets.

\subsubsection{Experiment setup}

We tested on 106 utterances from LibriSpeech-clean-100 from $27$ females speakers and $24$ males speakers. To assess our method, we created three different complementing utterances for each \textit{Original} utterance: (i) \textit{Generated} - our model generated output on the corresponding minimal pair word. (ii) \textit{Vocoder only} - applying only HiFi-GAN process on the original utterance. (iii) \textit{Smooth concat.} - we randomly selected a speaker of the same gender and with the target phoneme preferring utterances from the same target word. we combined the audio sections using SoX toolkit (provides a cross-fade at the join, and a wave similarity comparison to help determine the best place at which to make the concatenation). This method was designed to automatically simulate resemble the Str{\"o}mbergsson method and compare the result to our \textit{Generated} results. Our evaluation assessed the perceived correctness and quality of the generated audio by using Amazon mTurk crowd-source platform with native American English raters. Each rater was presented between 60-100 audio samples for each experiment.

\subsubsection{Minimal pair}

Table \ref{tab:minimal_pair_phoneme} presents the accuracy result on phoneme identity texts using similar testing to ABX testing with minimal-pair words. The test asked participants to classify a random audio, from \{\textit{Vocoder only, Generated, Smooth concat}\}, between the target word, the corresponding minimal pair word, a control option - an example that was irrelevant and very different from any valid value and ``none of the above.'' 
We assume that the cause for higher performance in the \textit{Smooth concat.} method compared to our \textit{Generated} method is that in the former contains the exact corrected phoneme while the generator produced an approximate one. 

\begin{table}[h!]
  \vspace{-0.1cm}
\small
\renewcommand{\arraystretch}{0.9}
  \centering
    \caption{Minimal-pair word correction evaluation by mTurk. The rows with $\rho^*$ indicate the accuracy of the correction with the target phoneme $\rho^*$ .} 
    \vspace{-0.2cm}
    \label{tab:minimal_pair_phoneme}
    \resizebox{0.47\textwidth}{!}{%
  \begin{tabular}{l| ccc}
    \toprule
    Audio type & \textit{Vocoder only} & \textit{Generated} & \textit{Smooth concat.}   \\
  \midrule
  accuracy & \!\text{96.0\%}& \!\text{76.4\%}& \!\textbf{83.7\%}\\
  switched & \!\text{2.2\%}& \!\text{20.8\%}& \!\text{10.9\%}\\
  none above & \!\text{1.7\%}& \!\text{2.8\%}& \!\textbf{5.4\%}\\
  \midrule
  $\rho^*=S$ & \!\text{94.2\%}& \!\text{64.6\%}& \!\textbf{95.7\%}\\
  $\rho^*=SH$ & \!\text{97.8\%}& \!\textbf{90.0\%}& \!\text{89.4\%}\\
  $\rho^*=W$ & \!\text{97.9\%}& \!\text{69.7\%}& \!\textbf{74.6\%}\\
  $\rho^*=R$ & \!\text{91.8\%}& \!\textbf{86.2\%}& \!\text{77.2\%}\\
    \bottomrule
  \end{tabular}}
  \vspace{-0.6cm}
\end{table}

\subsubsection{Speaker verification}

The duration of each utterance was $\sim 0.4-1.2$ seconds ($1-3$ words). While cutting a small word from audio mid sentences could decrease the quality of perception and recognition, adding the context tends to be problematic since context can create biases in our perception, e.g., ``chapter one'' vs. ``chapter run''. Consequently, we only took unbiased context, which mainly resulted in adding articles or conjunctions if possible and unnatural yet possible combinations, e.g., ``gave one'' $\to$ ``gave run,'' or ``when then ship'' $\to$ ``when the sip.'' 

In the experiment, first, raters listened to the random speech utterance, from \{\textit{Original, Vocoder only, Generated}\} and then to three speech utterances of a specific speaker (we created three unique $5$ seconds utterances for each speaker; those utterances were not used for any other tasks). Then, the raters were asked to determine whether the random utterance is from the same speaker as the other three utterances. We got accuracy rates of $74.1\%$ on \textit{Original}, $74.7\%$ on \textit{Vocoder only} and $71.0\%$ on the \textit{Generated}. We suspect that the main reason for the low values for all audio types is that the tested utterance is too short, and therefore did not contain enough information for the speaker to be accurately recognized. Another possible factor is that LibriSpeech is a book reading dataset in which the speakers produce different voices for characters to match the book, confusing the raters.

\subsubsection{Perceived quality evaluation}\label{sec:mos}

Mean Opinion Score (MOS) is a numerical measure of subjective human evaluation, often used to measure synthesized speech quality. Raters were asked to listen to a random speech utterance from any possible audio type and choose a score from $1-5$ with increments of $0.5$. The score reflects audio naturalness and quality. The mean and std MOS results are as follows: \textit{Original} $3.77 \pm 0.96$, $3.56 \pm 0.94$ on \textit{Vocoder only} and $3.37 \pm 0.96$ on the \textit{Generated}. participants were significantly more likely to prefer our generated version than \textit{Smooth concat.} which received a MOS score of $2.96 \pm 1.06$.  
The dataset we used (LibriSpeech-clean-100) is known for relatively very clean and high-quality recordings, which suggests the raters scored relatively low overall, as the results on the \textit{Original} indicate.

In the last experiment we tested Conditional MOS: In contrast to standard MOS, we used a pairwise comparison to verify the statistical significance of our results. Based on a given audio reference, raters were asked to choose a score for the second audio, given that the reference has a default score of $3.5$. The score reflects how natural and of high perceived quality the audio sounds. Table \ref{tab:phonemes_conditional_mos} shows the difference in mean MOS scores between the different audio types.


\subsubsection{SSD and ASD testing}

Emphasizing the model's capabilities and possible contribution, we tested our model on untrained Hebrew language sample recordings with /s/ and /sh/ as target phonemes. We focus on /s/ and /sh/ since those phonemes can be used on a model that was train on English and tested on Hebrew.

We selected 51 utterances from 4 children and one adult who suffers from impaired speech, and 18 utterances from 4 children with ASD. The experiment was an A/B Test: raters listened to two speech utterances, one \textit{Generated} and the \textit{Vocoder only} audio. The listener was tasked to select which recording sounded correct to a target word and more natural. Every assignment was sampled $12-16$ times. The results showed the participants classified the \textit{Generated} speech as the correct pronunciation and more natural audio by $68.1\%$ of the time for all groups, $72.8\%$ for the ASD group, and $66.6\%$ for regular SSD. We computed the statistical significance of our result. Based on a binomial test (N = 912), participants were significantly more likely to select the \textit{Generated} speech as correct than the \textit{Vocoder only} speech (the worst the p-value was $3.2e^{- 12}$).






\section{Conclusion}\label{sec:conclusion}


In this work, we present a first attempt to generated a correct production from incorrect one, while maintaining the speaker's voice. Results suggest that our proposed model has a slight advantages in subjective listening tests. Our future work will focus on improving the phoneme embedding by looking at self-supervised learning algorithms to improve representation.


\bibliographystyle{IEEEtran}
\bibliography{references,foreign_accent_conversion}

\end{document}